# Cybercrime as a Service: A Scoping Review

Ema Mauko

University College London, emilija.mauko.22@ucl.ac.uk

Shane D Johnson

University College London, shane.johnson@ucl.ac.uk

Enrico Mariconti

University College London, e.mariconti@ucl.ac.uk

Cloud computing has drastically altered the ways in which it is possible to deliver information technologies in a service-led structure, however, this has also been reflected in the cybercrime domain. Cybercrime as a Service is an economic model where a technically skilled actor offers a given cyberattack as an end-to-end service to non-technical actors who pay a subscription fee for said service. The services, which can vary in scope, targets, and delivery modes, include everything from the vulnerability discoveries, delivery of the attack, and the attack itself to financial rewards to the subscriber. In this scoping literature review, we analysed 195 articles from both academic and grey literature with a view of investigating the services articles studied, the methodological approach the how the CaaS model is predicted to develop in the future. Our review indicates that with further commercialisation of the model will further lower the barrier of entry to the cybercrime realm, increase sophistication of the attacks and increase resilience of the service providers and their ecosystem which will result in harder shutdowns of services by the authorities. Furthermore, as the model becomes more accessible, groups such as organised crime groups, extremist actors may use them as well, which may have implications for criminal activity in both cyber and physical domains.

# 1 INTRODUCTION

Recent developments in cloud computing and a global pandemic accelerated changes in our way of living; a life that is now more online than ever before. Although this new way of life has facilitated substantial improvements to everyday activities, it has also increased the range and volume of online criminal opportunities that bad actors can and do engage in. In addition to there being changes in criminal opportunities online – and perhaps in response to this – there has been a change in the way that many criminals operate. Where many offenders would previously have worked alone to commit every aspect of an online offence, a relatively new business model, referred to as Cybercrime as a Service (CaaS), has been evolving where some parts of a criminal offence are effectively outsourced.

Cybercrime as a Service (CaaS) refers to an economic model where a technically skilled actor offers a tool kit as a packaged service with easy access to tools and frameworks which provide all the services a less experienced bad actor needs to carry out a successful cyberattack that would otherwise likely be out of their reach [1-3]. CaaS is analogous to the Everything as a Service (XaaS) paradigm, which involves the creation of new services as an assembly of independent services available within the environment [4]. Most commonly, when discussing XaaS, the industry will talk about Software as a Service which has been defined as a "software delivery paradigm in which the software is hosted off-premises and delivered via the web" [5]. Services are available upon request, and the process operates purely as a business with order purchasing, deliveries, and invoicing [6]. In both legitimate and illegal businesses, this model also offers autonomous scaling for applications and has moved the payment system from pay-as-you-go (i.e. payment per action) to a subscription-based model of payment. A common model in the CaaS economy comprises three important groups of actors: (i) A (very) small group of highly technically skilled producers (operators) that provide the service of the crimeware exploit (e.g. ransomware, trojans, DDoS etc). (ii) A slightly larger group of actors who package, monetise, distribute, and advertise as well as provide support for the service. And lastly (iii) Buyers who purchase the service and use it for personal gains [6, 7]. The primary places where CaaS is advertised are hacker forums, hacker shops, and the dark web, however, the actual transaction and handover often happens over a private communication channel [8]. Consequently, much of the academic research on this topic involved analysing underground forums and markets where such advertisements can be found.

From a criminological perspective, one may consider the rational choice perspective which suggests that an offender is a rational actor who aims to maximise utility by evaluating the perceived risk, effort, and benefits of their actions [9, 10]. Applying this to the CaaS model, there is a possibility of high reward and low risk to both providers and users of the service. It reduces the barrier of entry, and hence effort, for less experienced actors to perform complex attacks. This then enables increasingly sophisticated attacks to be conducted which can fuel the rapid development of new advanced threats [11, 12]. Furthermore, CaaS can enable cybercriminal organisations to be more agile and structured not unlike an IT company with multiple roles in order to maximise the profit and increase the reach of services (see Figure 1). Moreover, there are indications that underground markets operate according to economic principles. For instance, the interaction of supply and demand serves as a valuable pricing mechanism for some services, tools, and goods related to cybercrime [13].

This paper presents a scoping review of the emerging literature on CaaS with a specific focus on what services are investigated in the literature, the methodological approaches used, and what the predicted future developments of CaaS are To our knowledge, there is no systematic nor scoping literature review (SLR) on the topic thus far. As CaaS encompasses a wide range of illicit activities facilitated through digital platforms, often involving complex interactions between technological, economic, legal, and sociological domains research may be conducted by those in a variety of disciplines. As such, an SLR would be of considerable value to take stock of what is a rapidly developing problem, but also to synthesise what is known across a fragmented literature disciplines. Moreover, and importantly, an SLR provides a rigorous and transparent methodology for evaluating the quality and scope of existing studies, which can highlight gaps in the literature, guiding future research (including the need for studies that employ more rigorous methodologies) and informing evidence-based policy and practice.

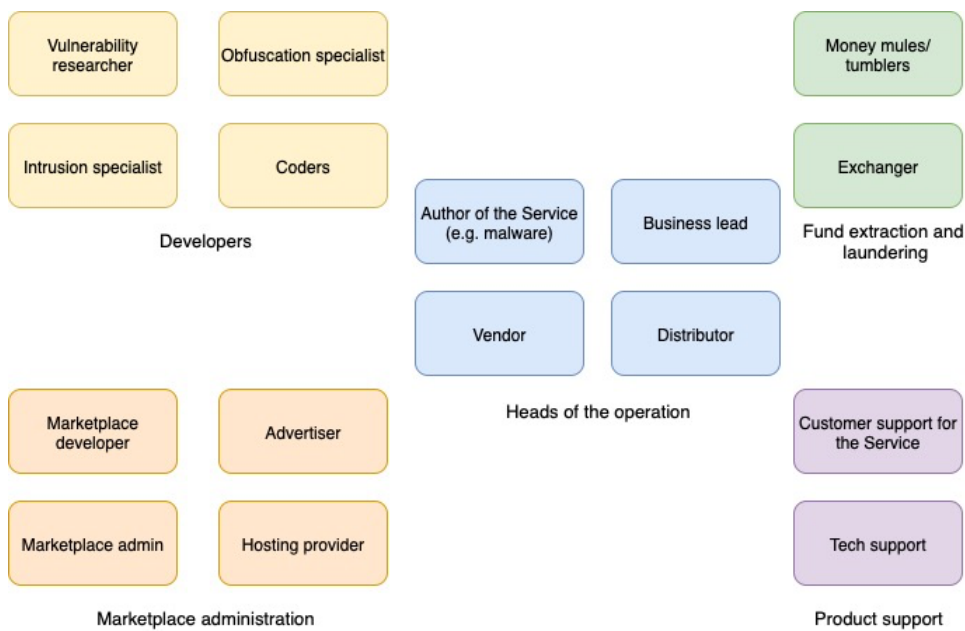

Figure 1: *Example roles of a CaaS operation[13-15]*

The remainder of the article is organized as follows: First, the reader is presented with a background section discussing the development of CaaS, CaaS in the present day and the challenges of studying the topic. Next, we describe the methodology adopted. The subsequent section summarises the articles included in the review and discusses the common themes identified in the papers apropos services analysed, future trends as well as the methodological approaches taken in the literature. Section four presents a discussion of the findings and their implications and outlines some of the limitations of the paper. Lastly, section five concludes with suggestions on how future work may shape the field of CaaS research.

## 2 BACKGROUND

### 2.1 Development of CaaS

Figure 2 provides an overview of some key developments in the timeline of the evolution of CaaS. Even though the term Crime as a Service, is relatively new, the idea of renting cybercrime tools has been around for a while, albeit for more specific and narrow use cases. There were numerous events and cyberattacks that used the CaaS-like economy, without referring to it as such. For example, botnet economies and phishing kits have been around since the early 2000s. Even then the kits, usually in the form of a website, contained a ready-to-deploy package, and usually also contained detailed usage instructions [16]. However, these instances would not necessarily be considered a "service" according to the contemporary definition which involves contracting an entity to do something for a period of time and not owning or purchasing a piece of software outright [17]. Furthermore, in the past, the buyer of the product still had to do quite a bit of work on their own to generate their own attack successfully. The same can be said about the malware toolkits that emerged in 2006 and were very popular in the 2010s [18-20]. In 2011, a source estimated that more than 60 % of malicious websites used DIY crimeware kits to do their dirty work [19]. However, again, these DIY malware toolkits did not come as a self-sufficient service. Rather, they were a set of instructions an actor had to assemble in their preferred way. That is, they provided an enabling technology but not an end solution [21]. The turning point for these kits was MPack– the first known crimeware kit sold for profit [22]. This was sold as a script and consequently, there were issues with kit piracy i.e., copying said script, changing it a bit and reselling it as a new product. From this point on, new crimeware kits often came with anti-piracy features, not unlike commercial software. This included limiting the number of installations possible on different domains and so on. This slowly gave rise to a more professional model of crimeware kits that can be thought of the as predecessors of CaaS as we know it today. Reflecting this, in the early 2010s, researchers began seeing trends such as cybercrime services offered with additional services such as hosting, configuration and the spreading of malware [23].

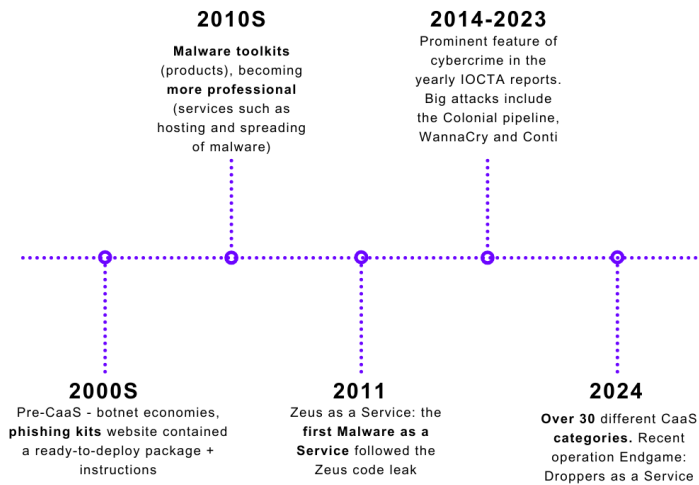

Figure 2: Timeline of CaaS development

Arguably, the first instance of a Malware-as-a-Service followed the leak of the Zeus 2.0.8.9. malware source code in 2011. This meant that the sale of the crimeware became unprofitable, leading to "Zeus as a Service" as an alternative monetisation strategy [24]. Zeus as a Service was then offered as a subscription service including hosting, configuration, and installs – much of what actors can now expect when purchasing a Malware as a Service. Since then, multiple other cybercrime attacks have started to become more industrialised, scalable to consumers, and offered as services. The CaaS model has been painted as a prominent feature of the cybercriminal underground in the yearly Internet Organised Crime Threat Assessment, assessing international cyber threats and published by Europol [2].

## 2.2 CaaS: the extent of the problem

As of 2025, CaaS has evolved into a dominant force in the cybercrime ecosystem, with some sources estimating it accounts for over 57% of all observed cyber threats [25]. The integration of offensive artificial intelligence, automation, and cloud-based infrastructure has enabled even low-skilled actors to launch sophisticated attacks with minimal effort [26]. These platforms now offer plug-and-play kits, customer support, and even performance guarantees, mirroring legitimate SaaS business models. Nowadays, over 30 different CaaS categories have been identified based on the type of service the frameworks promise [27]. These range from very targeted attacks, such as Hacking as a Service and Distributed Denial of Service as a service, to Ransomware and Data as a Service. However, there is no one complete list of services used in the literature and the services studied differ from one author to another. Together with no universally accepted definition of CaaS, there exists no accepted typology for services within the CaaS literature. Thus, research tends to lack an overview of what CaaS services are being offered and where in the CaaS ecosystem they are placed.

Authors also tend to disagree about which services are most prominent. For example, [27] have found that the 'Botnet as a Service', 'Reputation escalation as a Service' and 'Traffic as a Service' categories were the most prominent in terms of both supply and demand. However, in 2020 [28] reported that there had been an increasing amount of Malware as a Service, especially mobile malware, which was not as popular previously, whilst the most popular service with the largest impact was Ransomware as a Service (RaaS). It is estimated that as many as two-thirds of ransomware attacks are enabled by the RaaS model, although that may be difficult to determine purely based on the attack itself [29]. RaaS development is especially interesting as some of the latest trends include using supply-chain attacks to compromise the networks of large corporations and public institutions and utilising new multi-layered extortion methods and ransomware affiliate programs. Affiliate programs are offered as an alternative to a classic cloud-based subscription model that makes the service available to anyone who pays a subscription fee. Instead, levels of affiliates receive money as a commission from the ransom they extort from victims and give a percentage of that money to the developers of the attack in a similar way that investment schemes are run. These programs have become more sophisticated, offering tiered commission structures and performance-based rewards. This model has proven more scalable than traditional subscription-based services, allowing developers to profit from successful attacks without directly engaging in operations [30].

From a more infrastructural standpoint, grey infrastructure services such as bulletproof hosts (network infrastructure that is resilient to takedown attempts and complaints of illicit activities [31]), Virtual Private Networks (VPNs) and rogue, anonymous cryptocurrency exchanges such as Monero and swapping services, provide safe havens for these criminals. The market is expanding rapidly [8, 11], which requires it to be accessible to new entrants. This means that to operate effectively, the old and constrained model of closed, vetted, and trust-based criminal networks like those that used to be used on the darknet will be sub-optimal [27]. Completely open advertising carries increased risks though which could offset the benefits realised (i.e. reaching a wider user base). Consequently, although the markets and advertisements are more open in nature than they used to be, many offerings are only communicated directly to potential buyers via instant and private messaging. Thus, open markets represent merely a fraction of the CaaS landscape. Certain services are even more accessible. For example, Deepfakes as a Service have been shown to be easily accessible via apps such as DeepFaceLab, ZAO, and FaceSwap which are offered as mobile apps freely available on app stores without appropriate safeguards in place [32]. Some services may be also found on gig-based platforms where freelancers offer these services [33, 34]. CaaS services are increasingly partially or even fully automated, such as nudifying bots, monitoring social media feeds for real time vulnerability search, generation of phishing content, negation of ransom payouts and so on [35]. In some instances, e.g. nudifying bots on Telegram and mobile apps, the services are offered as "freemium" packages (basic settings are free, whilst some features require a payment "premium"). Paid services would include skipping long queues to generate a video or removing water-marks [36].

## 2.3 Challenges with studying CaaS

The CaaS-related threat space still requires a substantial amount of research; however, it is also important to note that when investigating CaaS itself, there are numerous issues to consider. Firstly, there is no clear and specific definition of what exactly an offer has to contain in order to be considered a CaaS. The existing literature on the subject uses a loose definition, such as that quoted at the start of this paper. Thus, one may not be able to properly define when exactly CaaS emerged and classify all the different services offered on the market. Without clear definitions, research in the field may not measure the same phenomenon and come to misunderstandings, which may affect the validity and generalizability of the findings. The second challenge to consider when studying CaaS is that accessing all relevant data may be tricky as a lot of exchanges happen on private channels (one-to-one communication via direct messaging) or dark web forums. These are much more difficult to find and reach, due to the dark web structure [37]. Researchers have aimed to counter the problem of accessing data by using web crawlers to collect data from underground forums. While this is representative of real-world forensic investigations, it is not systematic, and is hard to replicate [38]. The data may be skewed due to anti-crawling techniques, and the forums may not provide access to relevant data for the illegal activity in question. The third issue is that some markets and vendors use false categories to obscure the products they are offering. They may also use different words and wording such as homonyms to avoid detection [39]. Lastly, it is important to filter out fraudulent threads and posts – those that do not actually offer the services discussed – although these might be difficult to detect [40].

## 2.4 Motivation and aims for this project

To our knowledge, there is no previous existing systematic synthesis on the topic of CaaS. The aim of this research paper is to conduct a multidisciplinary scoping review of the literature on CaaS to take stock of what is known, identify potential research gaps and evaluate the extent of the research. To do this, the following research questions (RQ) were developed:

*RQ1: What types of services are discussed in the literature?*

As previously mentioned, there is no one established definition of CaaS and a list of services that would fall under this category, therefore, the first step in exploring what the literature has studied is to evaluate what services are being investigated.

*RQ2: What research methods and data have been used to research CaaS?*

In order to prioritise and evaluate the reliability of the findings in the literature and identify research gaps in the field, it is important to summarise the types of methodologies previous researchers have used to study CaaS.

*RQ3: What future trends of CaaS are discussed in the literature?*

Since CaaS has been an emerging trend in the literature and the news, it is valuable to synthesise what authors deem to be emerging trends and what direction the CaaS model may take in the future. This will be important in order to keep up with the arms race against malicious actors and CaaS providers [41].

To address the research aims, the academic and grey literature (see below) was reviewed using a systematic and explicit methodology to identify, select, and critically evaluate the findings of included studies. Scoping literature reviews are considered original work because they are conducted using rigorous methodological approaches. This aggregation of existing evidence to address specific research questions is then used to support the development of evidence-based guidelines for researchers and practitioners.

# 3 METHOD

This section outlines the aim of the research and describes the scoping review methodology employed– which follows the PRISMA statement guidelines [42] – including the specification of the inclusion criteria and search strategy used. It also details how data was extracted from the included articles and how the findings were synthesised. [42]. We used a content analysis approach as this is especially suited for the analysis of large, multidisciplinary datasets [43]. The full protocol explained in this section is available at the link below to allow full reproducibility of the study.

## 3.1 Scoping Review or "Methodology"

Prior to commencing the review, we developed a research protocol[1]. This detailed the research questions to be addressed, what electronic databases would be searched, what our inclusion criteria and piloting strategy were, how we would manage the literature, and how studies would be selected and what data would need to be gathered from the papers. All aspects of the protocol are expanded upon below.

*3.1.1 Electronic databases searched*

The following academic search engines were searched: Web of Science, IEEE Xplore, ProQuest, ACM Digital Library, Scopus in addition to Policy Commons, and BASE, which provide coverage of grey literature databases. Table 1 provides a summary of what these databases cover. This distinction of grey and academic literature based on the publishing venues (academic publishers for academia and non-academic publishers for grey literature) is followed throughout the remainder of the paper to distinguish the source of the articles [44].

Table 1: Search engines description

| Search engine | Databases description |
|---|---|
| Web of science | Conference Proceedings Citation Index, Science Citation Index Expanded, Social Sciences Citation Index, Arts & Humanities Citation Index, and Book Citation Index |
| IEEE Xplore | Indexed articles and papers on computer science, electrical engineering and electronics from the Institute of Electrical and Electronics Engineers (IEEE) and the Institution of Engineering and Technology |
| ProQuest | Library & Information Science Abstracts (LISA), Proquest Central (Criminal Justice Database, Computing Database, Library Science Database, Science Database, Social Science Database, Psychology Database and continent-specific databases covering technology and social sciences (such as Australia & New Zealand Database, Continental Europe Database, East & South Asia Database, East Europe & Central Europe database etc.), ProQuest Dissertations & Theses Global |
| ACM Digital Library | The Full-Text Collection of all ACM publications, including journals, conference proceedings, technical magazines, newsletters and books on the following topics: Artificial Intelligence, Machine Learning, Computer Vision, Natural Language Prcessing, Information Systems, Search, Information Retrieval, Database Systems, Data Mining, Data Science, Web, Mobile and Multimedia Technologies, Society and the Computing Profession, Applied Computing: Industry/Business, Physical Sciences, Life Sciences, Education, Law, Forensics, Arts/Humanities, Entertainment, Graphics and Computer-Aided Design, Networks and Communications, Architecture, Embedded Systems and Electronics, Robotics, Hardware, Power and Energy, Human Computer Interaction, Security and Privacy, Software Engineering and Programming Languages, Computational Theory, Algorithms and Mathematics |

---

[1] See the protocol here: https://liveuclac-my.sharepoint.com/:w:/g/personal/ucabea2_ucl_ac_uk/ER8j-2CRlJxNmyhr5A3TDAcBdPt9UIyxgZfqSa0SuwsLLQ?e=B61YHs

| Scopus | Elsevier's abstract and citation database - Content on Scopus comes from over 5,000 publishers and must be reviewed and selected by an independent Content Selection and Advisory Board (CSAB) to be, and continue to be, indexed on Scopus. |
|---|---|
| BASE | 300 million documents from more than 10,000 content providers[2]. Content includes the metadata of academically relevant resources – journals, institutional repositories, digital collections etc. |
| Policy Commons | Collection of research articles from policy experts, think tanks, inter-governmental organisations (IGOs) and non-governmental organisations (NGOs). |

We decided not to search for industry reports for two reasons. First, these are not systematically indexed in electronic search engines such as those above. And, more importantly, we felt that there was a risk that the industry reports might present a biased picture of those threats for which the companies (as authors) provided solutions.

### 3.2 Search strategy

The first stage of the study involved the use of a pre-defined and optimized search query to identify relevant articles. A preliminary search of the literature suggested that over 30 different services have been identified as forms of CaaS, spanning from very targeted attacks, such as Hacking as a Service and Distributed Denial of Service as a service to Ransomware and Botnet as a Service. To capture as many malicious services across the interdisciplinary research community, the search terms used were not restrictive and were designed to capture as many malicious services as possible. The terms used were:

```
"*ware" NOT "software" NOT "hardware" NOT "aware" OR
"crime" OR "cybercrime" OR "attacks" OR "botnet" OR "mal*"
                        AND
       "as a service" OR "as-a-Service" OR "for hire"
```

NOTE: "*" denotes a wildcard, such that for example "*ware" would identify "ransomware", "crimeware", "malware" and so on.

### 3.3 Inclusion criteria

A two-stage screening process was used for most articles (see below). Initially, the titles and abstracts of all articles identified were screened using our inclusion criteria (subsequently the full texts were reviewed). For articles to be included the following criteria had to be satisfied:
- Studies must discuss an "as a Service" model for malicious purposes online (e.g. a study focused on the use of machine learning as a Service for legitimate purposes would be excluded).
- For the academic literature, the abstract had to mention the "as a Service" model of a service used for malicious purposes (e.g. a paper talking about "Ransomware as a Service" would be included, whilst an abstract discussion about "Drones as a Service" used for grocery delivery would not).

---
[2] See the content provider index here: https://www.base-search.net/about/en/about_sources_date.php

- As the majority of articles in the grey literature consisted of longer reports often spanning over 50 pages, abstracts were not good indicators of what is included in each article. As with other scoping reviews (e.g. , therefore, these articles were screened in full and the keywords "as a Service" were searched for to determine whether the article should be included or not.
- Papers must mention the as a Service model or similar terminology (e.g. for hire) more than once excluding the abstract. This was to eliminate articles that only mentioned the model in passing.
- Articles had to be published between January 2017 and the date of our search (December 2024). 2017 was chosen as it represented a turning point for CaaS and research on it with the takedowns of large markets such as Hansa and Alphabay which provided a lot of insight into the cybercrime ecosystem and its operation [45]. Furthermore, the Mirai botnet code was publicly released towards the end of 2016, and this was associated with an increase in botnet and further malware as a Service products on the market [46]. Moreover, the research questions are about recent trends in the CaaS ecosystem. As such, we believe that the 8 years' worth of research included here ensure that the review captured contemporary as opposed to historic trends.
- Articles had to be available in English.

Some of the articles identified discussed "as a Service" technologies that potentially could be misused. For example, some studies discussed dual-use technology services, such as Drones as a Service or Machine Learning as a Service. In such cases, papers were only included if the malicious use of the service was explicitly discussed by the authors of the papers concerned.

### 3.4 Selection of studies

After completing the search, identified citations were imported into Rayyan[3], a management tool for systematic analysis. First, we identified and removed duplicate entries. Next, the abstracts of 300 randomly selected articles were screened by at least two researchers, with 50 articles screened by all three to ensure that the inclusion criteria were clear and could be applied reliably. In the case that the inclusion could not be determined based on the abstract alone, we screened the full text. Articles were coded as included, excluded, or "maybe" by each author. When the code was "maybe" all three reviewers read the abstract and discussed its inclusion. To quantify inter-rater reliability, we calculated the prevalence- and bias-adjusted kappa (PABAK) statistic. The score of 0.86 indicated a high degree of agreement amongst the researchers.

All titles and abstracts were then screened by the lead reviewer. Any articles that were categorised as "maybe" were then discussed by all three researchers and a decision made as to whether to include them. Articles were then screened on the basis of their full text.

### 3.5 Data extraction and management

Different studies used different methods to study CaaS. As discussed elsewhere (e.g. [47]), when reviewing the literature, it is important to explicitly categorise the type of evidence produced for at least two reasons. First, taking stock of the methods used is important for researchers to see where there are gaps in the literature. And second, different methods have different strengths and weaknesses. For example, some approaches enable researchers to predict what might happen in the future, which is useful for anticipating new threats, but do not (for example) facilitate the estimation of the prevalence of the use of a particular CaaS today. Other methods may assist with the latter but provide little insight into the former.

---

[3] https://www.rayyan.ai

The aim of the project was to explore the extent of the research that has been completed regarding CaaS. To achieve this, the inclusion criteria captured articles with different levels of importance to the topic of CaaS. There are at least two important dimensions: 1) studies can vary in terms of the methodology employed and the data analysed, which can influence the quality of the evidence; and, 2) studies can vary in the extent to which they focus on CaaS, with it (for example) being the primary focus of enquiry or of peripheral interest. Figure 3 visualises this and identifies four categories of evidence (I-IV) for studies that involve an explicit methodology. These four categories are defined in Table 2, along with a fifth category, which is reserved for studies that do not include an explicit methodology that would facilitate their replication by others.

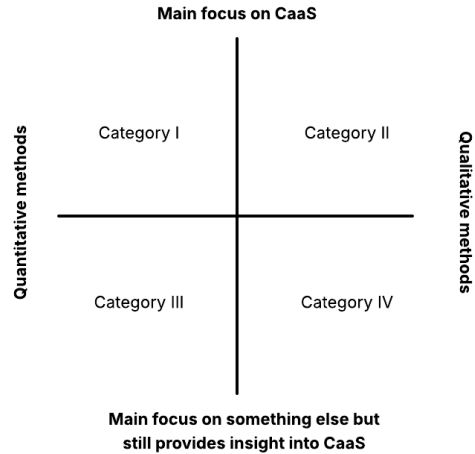

Figure 3: Categorisation of evidence (for studies with an explicit methodology)

Table 2: Categorisation of evidence description

| Evidence type | Description | Label |
| --- | --- | --- |
| Quantitative analysis of as a Service attacks | Paper uses (primary or secondary) quantitative data to investigate a Cybercrime as a Service operation. E.g., a paper provides a deep dive into Conti ransomware strands and reverse engineers the technical elements of this known RaaS operation. | I |
| Qualitative analysis of as a Service attacks | Paper qualitatively analyses real world data with regards to the CaaS model. This evidence type includes realistic and thematic analyses, systematic literature reviews (SLR) and theoretical papers in the field of mathematics and economics. SLRs are included in this category as they employ reproducible methods, have a sample of data which is analysed using thematic analysis and thus more align with the qualitative category rather than (speculative ad-hoc) literature reviews. E.g., a case study that includes five CaaS cases where the methodological approach is a thematic exploratory analysis that uses police files, formal and informal interviews and court transcripts. | II |
| Quantitative study of attacks that **can** be offered as-a-service model | Paper quantitatively analyses a cybercrime operation that is not limited to the CaaS model. The paper briefly discusses the cybercrime activity and its implication as part of the CaaS ecosystem. Such findings are relevant, but CaaS is not the main topic of the paper. Example papers include those for which there is a comparative analysis of ransomware detection algorithms, for which some are distributed as part of a RaaS model. | III |
| Qualitative study of attacks that **can** be offered as-a-Service model | Paper qualitatively analyses cybercrime operations of which datapoints may not be limited to the CaaS model. The paper briefly discusses the cybercrime activity and its implication as part of the CaaS ecosystem. Such findings are relevant, but CaaS is not the main topic of the paper. An example would be a paper thematically analysing recent case studies of ransomware in the news. In such papers, the majority of findings are focussed on the dissemination of ransomware and the victims, and while RaaS is discussed, this is only brief (e.g. this may be limited to a single sentence). | IV |
| Literature review | This category includes ad-hoc literature reviews (not systematic literature reviews) and essays. No primary data is analysed in such studies. Data may be supported by relevant case studies; however, these are only brief overviews. | V |

### 3.6 Data analysis: Content Analysis

Following the sifting of articles, the full text analysis commenced. The analysis strategy adopted was content analysis which is a widely used method for identifying, analysing, and interpreting patterns of meaning within qualitative data [48]. In particular we used deductive (or directed) approach to qualitative content analysis which is employed when the investigation is informed by extant literature, theoretical frameworks, or established conceptual models [49-51]. It involves systematically coding data and then categorising these codes into broader themes that represent significant concepts related to the research question. We used a structured analysis matrix derived from research questions and piloted with the first 20 papers and then remained fixed for the rest of the analysis, representing a clear top-down deductive logic. Nevertheless, should specific textual segments elude these predetermined classifications, content analysis allows for the researchers to adopt an inductive secondary process to allow new categories to emerge directly from the data [49]. While this introduces an element of inductive reasoning, it remains subordinate to the primary deductive orientation of the study [52]. The themes in this paper included the topic (main aim of the paper), what services the paper spoke about (all services the paper focussed on were identified), the specific context of service (if not broad – what environment was the paper focussing on – e.g. business security, social media…), method of the paper (qualitative/ quantitative/ mixed, data used, and specific methodological approach used), categorisation of evidence (see table 2), past trends (the main historic CaaS trends the paper identified), future trends (if present, what predications of CaaS the authors discuss – this could be emerging victims, techniques, vulnerable infrastructure, enablers…), implications (e.g. what does that mean for society, victims, offenders, and response), and any additional comments (this could be anything that would enrich the discussion of the topic that was not part of the research questions). The results below represent a synthesis of this coding.

## 4 RESULTS

### 4.1 Summary of search results

The initial search yielded a total of 3,896 published articles that were screened for eligibility (see the PRISMA chart, Figure 4). After the removal of duplicates, we screened 2,151 abstracts based on the inclusion criteria described above. Of these, 386 articles were included for full-text screening. At this stage, a further 134 articles were excluded due to them not meeting the inclusion criteria such as not mentioning CaaS anywhere in the paper but the abstract. The reported findings are thus based on 215 articles, of which 103 were academic articles and 112 were articles from the grey literature.

The articles from the grey literature were mostly reports produced by a variety of organisations, with the most being published by the European Union (44) and its various research bodies such as the European Parliamentary Research Service, European Union Agency for Asylum, European Monitoring Centre for Drugs and Drug Addiction and the European Economic and Social Committee. Some of the grey literature was published by international organisations such as the World Economic Forum (3), the Centre for Global Cooperation (1) and the United Nations (3). One article was produced by the African Union. The remainder consisted of reports produced by think tanks or governmental bodies of the following countries: Canada (3), Australia (3), India (12), Singapore (1), Ireland (1), Brazil (1), Switzerland (4), Sweden (1) the United Kingdom (6) and the United States (22). As shown in Figure 5, the number of articles from both the academic and grey literature was relatively evenly distributed across the first four years but saw an increase from 2022 onwards (the figures for 2024 are for six months).

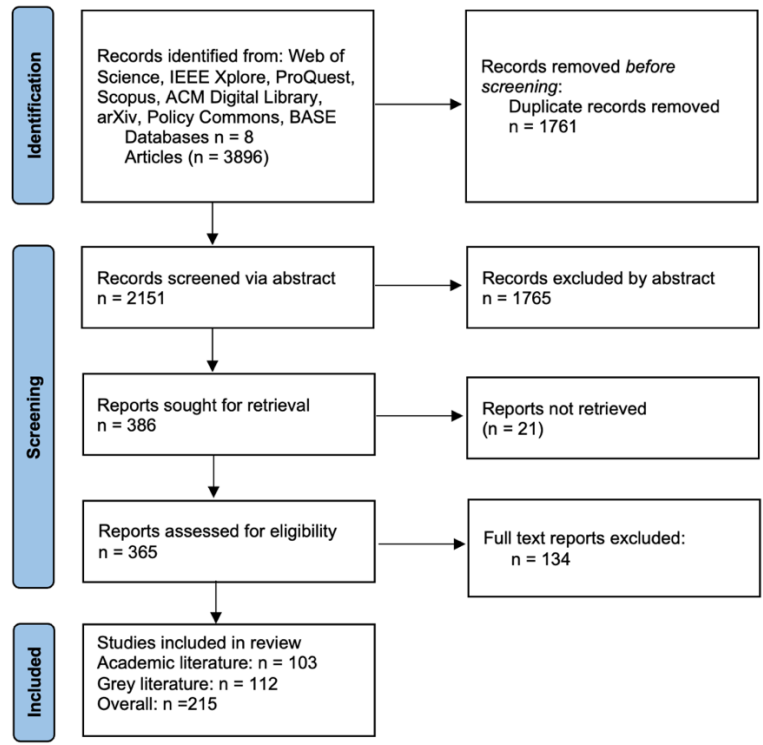

Figure 4: PRISMA chart denoting article selection process

## 4.2 Services discussed in the literature

The first research question concerned the types of services discussed in the literature. Tables 3 and 4 summarise what services were discussed in each paper in the academic and grey literatures, respectively. Where a paper discussed more than one service in detail, it was counted as doing so accordingly. As such, the cumulative count of services shown in Tables 3 and 4 exceed the number of papers included in our sample. For the academic literature, papers are also categorised according to the type of method used and the type of evidence generated (see Table 2). The articles were therefore split based on whether they specifically focused purely on CaaS ecosystem or services (Categories I, II and V) or not (Categories

III and IV) and whether they deployed quantitative (I and III), qualitative (II and IV) methodologies or consist of literature reviews (Category V). Grey literature publications did not report the methods used – all being written in a narrative report format – and so such a breakdown was not necessary or possible. This is discussed in more detail in a later subsection.

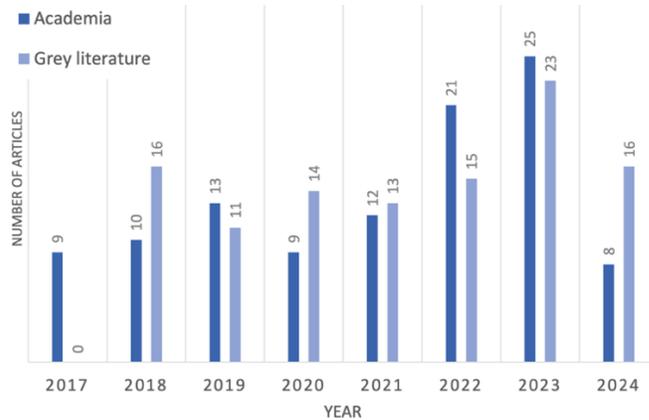

Figure 5: Number of articles per year

*4.2.1 Services discussed in the academic literature*

As shown in Table 3, in the academic literature, the most commonly discussed service was ransomware, which featured in 40% of the academic papers. The second most common topic was the introduction of Cybercrime as a Service in general (discussed in approximately 20% of papers), in which authors of papers either introduced the CaaS model and its implications as a prelude to introducing other services or just studied the model and its ecosystem itself in depth. Overall, the vast majority of crimes studied in the literature were cyber dependent. For example, many papers studied established cybercrime attacks such as malware (malware in these cases did not contain ransomware, i.e. no ransom was requested), botnets and Distributed Denial of Service (DDoS) attacks, and a smaller number of papers presented some of the more modern attack vectors such as concession abuse and Propaganda as a Service. Services such as phishing, residential proxies, pay-per-install, money laundering and Obfuscation as a Service are interesting, as they can be used by themselves, but are often used in tandem with other malicious activities. For example, an individual may wish to use their own ransomware attack. However, they may not have a mode of distribution and may want the ransom to be paid in a more discrete manner. In such cases, they may use Phishing as a Service and Money laundering as a Service to accomplish this. Similar overlaps can be observed for studies of Malware as a Service itself and its many subgroups. In addition to ransomware, which was discussed separately, stealers [53], adware [54] and spyware [14, 54] were all spoken about separately. The papers studied here rarely elaborated on the interconnectedness of the various CaaS nomenclatures or the interconnectedness and non-standardisation of services in the literature and in the wild.

*4.2.2 Services discussed in grey literature*

Table 4 shows the topics covered in the grey literature. For these studies, as discussed, we do not show the categorisation of evidence as the papers did not have methodology sections and hence could not be categorised. For this literature, 58% of the sample discussed CaaS in general. This was discussed either to just explain the CaaS model as a way for malicious actors to operate, or to introduce what CaaS is before discussing particular services (e.g. phishing, money laundering, malware etc) in detail. The second most discussed service was ransomware (47% of papers), followed by malware in general. As with the academic literature, numerous articles discuss more "traditional" cybercrime vectors of attack such as DDoS, Botnet, Phishing and Fraud as a Service. However, there was also discussion of more novel services such as artificial intelligence (AI) [28], swatting, doxing and mass surveillance as a Service that were not discussed in the academic literature. AI as a service was particularly discussed as being used for the creation of deepfakes and similar applications.

In particular, services that were found in in grey literature but not in academic literature included the following: Fraud (person or group of people dishonestly and deliberately deceiving a victim for personal gain of property or money or causing loss or risk of loss to another), Access (services specializing in acquiring access to target's systems), Piracy (unauthorised copying, distribution, or use of digital content), Injection attacks (malicious code is inserted into a system to manipulate its operations or access unauthorized data), Spoofing (technique where fraudsters fake caller IDs to gain victims' trust), Swatting, Doxing (Swatting refers to making hoax calls to emergency services and sending law enforcement to one's home unnecessarily, whereas doxing refers to unconsented release of someone's personal and identifiable information online. Both are considered as tactics for harassment), Sniffers and skimmers (Sniffers are tools that capture network traffic to intercept data, while skimmers are devices that steal payment card information during transactions), Mass surveillance / spyware (malware that collects user data without their knowledge or consent), and Cyber-enabled crime (migrant smuggling, document forging).

As described in the literature section of the paper, the larger CaaS ecosystem comprises many subparts. Recognising this, services such as Bulletproof hosting provider services [55] and Proxy services [56, 57] were mentioned as facilitators to the ecosystem in the grey literature. Furthermore, crimes that enable other offences such as Obfuscation, Identity theft, Access, Hacking, Sniffers and Skimmers and Money Laundering as a Service were also discussed in the grey literature. The latter were often discussed in the context of Organised crime groups and other larger malicious groups such as hacktivists. In addition, themes that emerged in the grey literature included cyber-enabled crimes such as migrant smuggling, document forging and acts of violence. While these offences primarily happen in the physical realm, they do also have a cyber domain [58]. For example, there are CaaS services for the creation of fake documentation (both online and physical passports), platforms with Services regarding intelligence sharing for migrant smuggling and platforms where physical violence services are advertised, including Murder as a Service.

### 4.3 The methods the literature employed to research CaaS

The second research question evaluated in this paper concerns the methodological approach authors employed in their research. As mentioned previously, the grey literature sample consisted of official reports, all of which presented ad-hoc reviews of the literature based on the authors interpretation of it. In some cases, the authors of a paper would illustrate their conclusions using data collected by their organisation or from case studies, however, these reports contained no methodology sections that would enable their replication or an objective assessment of the adequacy of the data or methods used. As such, this literature was more narrative and subjective in natureand the lack of methodological detail precluded us from categorising the evidence further.

As shown in Table 3, numerous methodological approaches were undertaken in the academic papers. In addition to there being variety in terms of the general approaches employed, there was considerable variation within each category of evidence. Considering the first category alone, quantitative analysis included a deep learning model for the detection of RaaS attacks using an intrusion detection dataset [59], the reverse engineering of real-world attacks by analysing leaked Conti ransomware source code [60] or using disassembler software that analyses binary code [61, 62], and even authors developing their own examples of ransomware and phishing operations as a proof of concept [63, 64]. Of course, the methodology employed in papers is influenced by the main theme of the paper.

Table 3: Services as reported in the academic literature and their categorisation of evidence. Different services (e.g. pay per install and exploits) have been analysed together or terminology has been used interchangeably by the authors of the previous work. The counts of papers are included in the final column with a number of empirical studies (non literature reviews) in the bracket.

| Service | Brief description | Categorisation of evidence | | | | | Count (empirical studies) |
|---|---|---|---|---|---|---|---|
| | | I (quant, CaaS focus) | II (qual, CaaS focus) | III (quant, not CaaS specific) | IV (qual, not CaaS specific) | V (lit reviews) | |
| Ransomware | Type of malware that locks victim's data or access to victims' system until the victim pays ransom. | [60], [61], [64], [65], [66], [59], [67], [68] | [62], [69], [15], [70], [71], [72], [73] | [74], [75], [76], [77], [78], [79], [80], [81], [82], [83], [84], [68], [85], [86], [87], [88], [89], [90], [91] | | [92], [93], [94], [14], [95], [96], [97] | 41 (34) |
| CaaS in general | The papers talk about CaaS model itself. | [37], [65], [98] | [30], [99], [70], [100] | [101], [102], [103], [104] | | [105], [106], [107], [108], [54], [109] | 17 (11) |
| Malware | Malicious software. | [110], [111], [64], [112], [113], [114] | [115], [100] | [116], [117], [118], [119], [120] | [114] | [121], [107], [54], [14], [95] | 18 (13) |
| Distributed Denial of Service (DDoS) | Attacker floods a server with traffic thus preventing the user from accessing their system. Referred to also as "booters". | [122], [123], [111], [124], [125], [126], [127] | [70] | [128], [129] | | [107], [108], [130] | 13 (10) |
| Botnet | Network of computers infected by malware that are under the control under the attacker and can be used for malicious purposes. | [124], [64], [98] | | [131], [128] | | [132], [107] | 7 (5) |
| Phishing | Form of social engineering where a scammer tries to deceive the victim to reveal their personal information or instal malware. | [63] | | [133], [134], [135] | | [136], [108] | 6 (4) |

| Service | Brief description | Categorisation of evidence | | | | | Count (empirical studies) |
|---|---|---|---|---|---|---|---|
| | | I (quant, CaaS focus) | II (qual, CaaS focus) | III (quant, not CaaS specific) | IV (qual, not CaaS specific) | V (lit reviews) | |
| Pay per install, exploits | Methods of delivering malware. The providers of these services distribute the malicious files supplied by their customers and get paid according to the number of downloads. | [98], [113], [114] | [30] | | | [121], [108] | 6 (4) |
| Money laundering | Concealing the source and destination of funds. | [68] | [137], [100] | | | | 3 (3) |
| Spyware | Subtype of malware that aims to gather information about a user without user's knowledge. | | | | | [54], [14] | 2 (0) |
| Residential proxy | A provider utilizes the hosts within networks to relay their customers' traffic, in an attempt to avoid server-side blocking and detection. These services often play prominent roles in the underground business world and provide further obfuscation of the malicious actors [138]. | [138] | | [139] | | | 2 (2) |
| Stealer | Specialized commodity malware that harvests credentials from infected hosts. | [53] | | | | | 1 (1) |
| Obfuscation | Services that conceal malicious content of a file to evade anti-virus detection [102]. | [102] | | | | | 1 (1) |
| Blackmail | Demand for payment or other benefit from someone in return for not revealing damaging information about them. | [140] | | | | | 1(1) |
| Propaganda | models "spinning" their outputs to support an adversary-chosen sentiment or point of view thus creating biased speech. | [141] | | | | | 1 (1) |
| Concession Abuse | "scam service in underground forums that defrauds online retailers through the systematic abuse of their return policies (via social engineering) and the exploitation of loopholes in company protocols" [142]. | [142] | | | | | 1 (1) |
| Hacking | Gaining unauthorized access to personal or organizational data through exploiting vulnerabilities in computer or network system. | | [40], [100], [143] | | | [106], [109] | 5 (3) |
| Impersonation, Identity theft | Attacker pretends to be someone else by stealing and using user profile credentials and other metadata to circumvent authentication systems. | [144] | | | | | 1 (1) |
| Number of unique articles for column | | 30 | 15 | 35 | 1 | 21 | |

To illustrate, consider Figure 6 which shows a Sankey diagram – a type of diagram used to quantitatively visualise the flow of materials or themes through networks and processes, whilst demonstrating their relationships and their transformation [145]. This is done by representing directed, weighted graphs, nodes of which in our case represent the split between grey and academic literature, methodology groupings and the main themes of given papers. It can be seen that papers study various RaaS topics – including state and organised crime group actors using RaaS, deep dives into the RaaS ecosystem, and RaaS detection, via various methodologies.

Table 4: Services as reported in the grey literature

| Service | Papers | Count |
|---|---|---|
| Cybercrime as a Service in general | [146], [147], [148], [149], [150], [151], [152], [153], [154], [155], [156], [157], [158], [159], [160], [161], [162], [163], [164], [165], [166], [167], [168], [169], [170], [171], [172], [173], [174], [175], [176], [177], [178], [179], [180], [181], [182], [183], [184], [185], [186], [187], [28], [188], [189], [190], [191], [55], [192], [193], [194], [195], [196], [197], [198], [199], [200], [201], [202], [203], [185], [204], [205], [58], [206] | **65** |
| Ransomware | [148], [157], [207], [208], [209], [210], [211], [212], [213], [214], [215], [216], [217], [218], [219], [220], [221], [222], [170], [223], [177], [182], [183], [224], [225], [185], [186], [28], [189], [191], [226], [194], [197], [199], [200], [208], [227], [185], [216], [204], [228], [229], [205], [230], [231], [232], [233], [234], [214], [56], [57], [206] | **53** |
| Malware | [153], [155], [214], [235], [167], [236], [225], [28], [189], [191], [237], [238], [194], [198], [214], [239], [56], [57], [240] | **19** |
| Money Laundering | [151], [162], [241], [225], [189], [190], [191], [196], [203], [58] | **10** |
| DDoS | [148], [213], [177], [181], [184], [236], [186], [191], [198] | **9** |
| Hacking | [242], [185], [232], [197], [208], [185], [56], [57] | **8** |
| Phishing | [28], [189], [190], [191], [198], [205], [233] | **7** |
| Fraud | [162], [181], [192], [196] | **4** |
| Botnet | [177], [182], [198] | **3** |
| Access | [28], [205], [206] | **3** |
| Impersonation, Identity theft | [150], [191] | **2** |
| Pay per install, exploits | [182], [240] | **2** |
| Piracy | [195], [201] | **2** |
| Injection attacks | [193] | **1** |
| Spoofing | [192] | **1** |
| Swatting | [184] | **1** |
| Doxing | [184] | **1** |
| Obfuscation | [190] | **1** |
| Sniffers and skimmers | [191] | **1** |
| Mass surveillance, spyware | [237] | **1** |
| Stolen data | [206] | **1** |
| Cyber-enabled crime (e.g. document forging) | [188], [167], [243], [168] | **4** |

Many papers employed a quantitative analysis (Category I and III studies), often experimental methodology to evaluate detection systems intended to protect against ransomware (see Figure 6). Quite commonly, this would entail of randomised and variable manipulation, machine learning approaches and would be done in a simulated environment. Some of those studies focused exclusively on RaaS samples, however, most studies evaluated a mixture of ransomware samples, only introducing the "as a Service" dimension of the threat in the introduction or discussion of the papers. Moreover, some of the main themes are correlated. For example, reverse-engineering a RaaS attack with a quantitative methodology can both inform the development of a detection system and help better understand the RaaS ecosystem model.

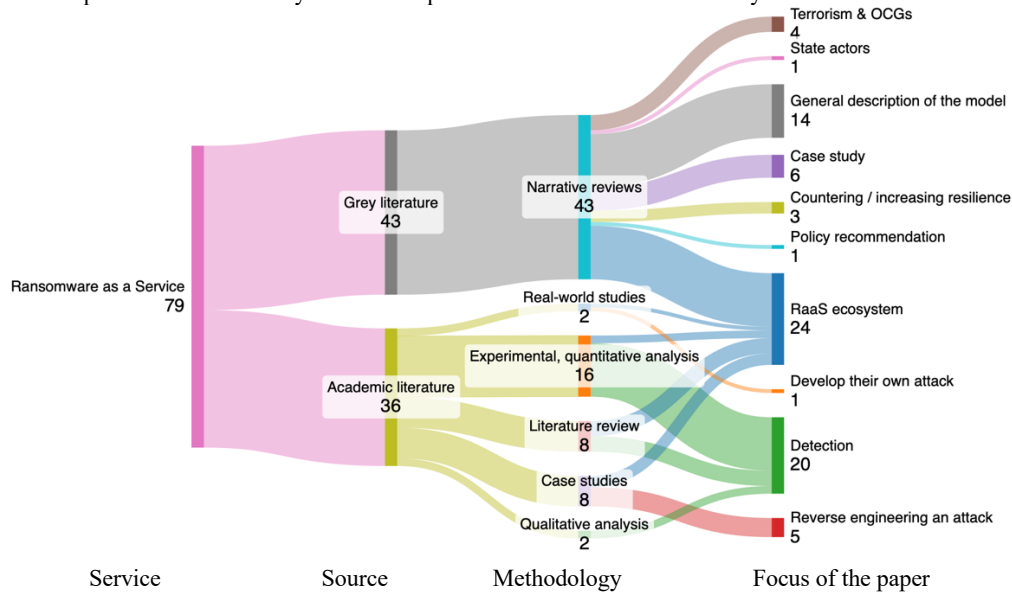

Figure 6: Sankey diagram demonstrating the methodology used in papers and the main focus of the paper

Compared to the more narrative approach taken in the grey literature, qualitative methods and ad-hoc literature reviews were used less often in the academic literature. On the other hand, when considering academic literature, qualitative methods and literature reviews were used less frequently. However, when these methods were used, they were used mostly to study the RaaS ecosystem, the topic most commonly discussed across the two literature samples. Apropos RaaS (and CaaS in general), discussions in the grey literature were mostly limited to their discussion as an emerging threat and as a challenge to law enforcement, sometimes discussing examples of larger RaaS attacks such as the Colonial Pipeline and its consequences.

The reviewed literature has examined a diverse range of data, which may be broadly categorised into primary data where researchers collected their own data from the "wild" and secondary, pre-existing data sources. A substantial proportion of the studies, approximately two-thirds (not including literature reviews), constructed datasets through active experimentation, sandbox execution, web crawling, and controlled cyber operations. These primary data sources included dynamic behavioural traces such as API calls, registry modifications, file system alterations, and hardware performance counter statistics, typically captured within sandboxed environments. Static analysis was also prevalent, particularly in category III, where researchers examined properties such as Portable Executable files to extract opcode histograms (to enable the analysis of a programs behaviour), metadata, entropy values, and so on. Network traffic data constituted another significant category, where several studies analysed packet capture files to extract conversation-level features such as IP addresses, port numbers, and byte counts. Some authors monitored malware download URLs over extended periods using high-interaction web crawlers to assess site activity and longevity. Controlled attacks were also used to generate traffic logs, including the purchase of DDoS services to target test servers, or the creation of synthetic victim accounts to examine hack-for-hire behaviour. In addition, researchers harvested data from online forums such as Hackforums.net, capture private chat logs from Pay-Per-Install services.

In contrast, some studies relied predominantly on secondary data sources, including public repositories, leaked materials, industry reports, and prior academic literature. For category III, malware samples were frequently obtained from platforms such as VirusShare and the RISS open-source collection, while dynamic logs were sourced from external sandbox services. Some authors analysed leaked source code and internal communications, such as the Conti ransomware binaries and chat logs, to understand operational strategies and development practices. Network datasets such as ransomware trackers were used for intrusion detection and traffic analysis. For papers that employed qualitative methods (category II and IV), the data ranged from police reports, netnographic observations from cybercrime forums, and insights from industry publications.

Several theoretical and conceptual frameworks were used to guide research. As CaaS in itself is a sociotechnical problem the reviewed academic literature stems from different fields often merging traditional criminological theories with economic and computer science perspectives. Despite this, numerous theories and conceptual frameworks were commonly employed by the researchers. First, the Design Science Research paradigm was frequently employed to construct and evaluate artefacts such as classification models and data-analytics systems, particularly in more quantitative papers. On the other hand, Routine Activity Theory [244] served as a foundational lens for interpreting the CaaS ecosystem, from a criminological perspective. Several studies also incorporate the Rational Choice [245] perspective to model attacker and victim decision-making processes, particularly in relation to ransom demands and payment behaviours. The CaaS business model was central to conceptualising cybercrime as a structured economy, with specialisation interpreted through analogies to capitalist labour divisions and entrepreneurship theory, highlighting affiliate alienation and market entry barriers. Legal and economic analyses invoked Opportunity Theory [246], Social Network Theory [247], and Joint Commission Theory to explain how offenders acquire expertise and form collaborative networks. Behavioural and detection-focused research relied on various attack-phase taxonomies and multi-level extortion models to structure their findings, be that threat classification or various unsupervised anomaly detection.

**4.4 Future trends**

The final research question concerned how the CaaS model might develop in the future. 53 % of the articles included in the SLR discussed future trends to some extent so the following section is based on those articles alone. In order to demonstrate which trends related to each given CaaS service and their sources, a Sankey diagram was created (Figure 7). The graphic first splits the literature that has identified any future trends into academic and grey literature, and then further into the predicted future trends that are further described in Table 5 and the service(s) that those trends would apply to.

Table 5 provides a summary of the future trends identified in both the academic and grey literatures. According to the literature, CaaS will play a prominent role in the future of cybercrime as cooperation among criminal elements and hostile entities evolves over time. The consensus across studies is that the CaaS ecosystem will expand in the coming years, both in terms of the ecosystem itself, and the range of services offered as more crime types move to a service-based model to optimise operations and maximise profits. Furthermore, some of the criminals operating the CaaS model already treat their activities as a full-time job [53], and most of the studies reviewed in the SLR predict that CaaS will continue to grow, becoming more sophisticated, and commercially viable. This, more formal and corporate structure may be done with the increase of subscriptions and the licensing of products, and the addition of even more specific roles into the ecosystem such as intrusion specialists, access brokers, and marketing strategists. This is especially discussed in one paper that looked at the chat logs of a large RaaS operation. In this paper, the authors highlighted the importance of the non-tech talks within the operation, and a focus on managerial/business-oriented roles and discussions [200]. Building on this, in the future, for example, more ransomware operations may continue to operate as RaaS, especially with the further professionalisation of roles such as access brokering and perhaps mergers of RaaS partnerships [227]. Similarly, the online fraud market is growing with many "as a Service" components being offered as facilitators and enablers, thus increasing the impact of criminal operations [192]. Further commercialisation and a move towards a more sophisticated business model may result in operators introducing new tools that significantly lower the entry barrier for less experienced/technical cybercriminals, therefore increasing its reach. With the expansion of the model, CaaS may encompass even more illicit activities in the cyber, physical, and cyber-physical domains. Some authors [13, 30, 248] suggest that as a consequence, the market may become saturated which would lead to decreases in the price of services, which in turn may mean that only the most resilient businesses, offering quality products will remain. Due to this business competition, it is anticipated that attacks will develop faster and be stealthier in order to provide better services and keep a competitive advantage [88, 94].

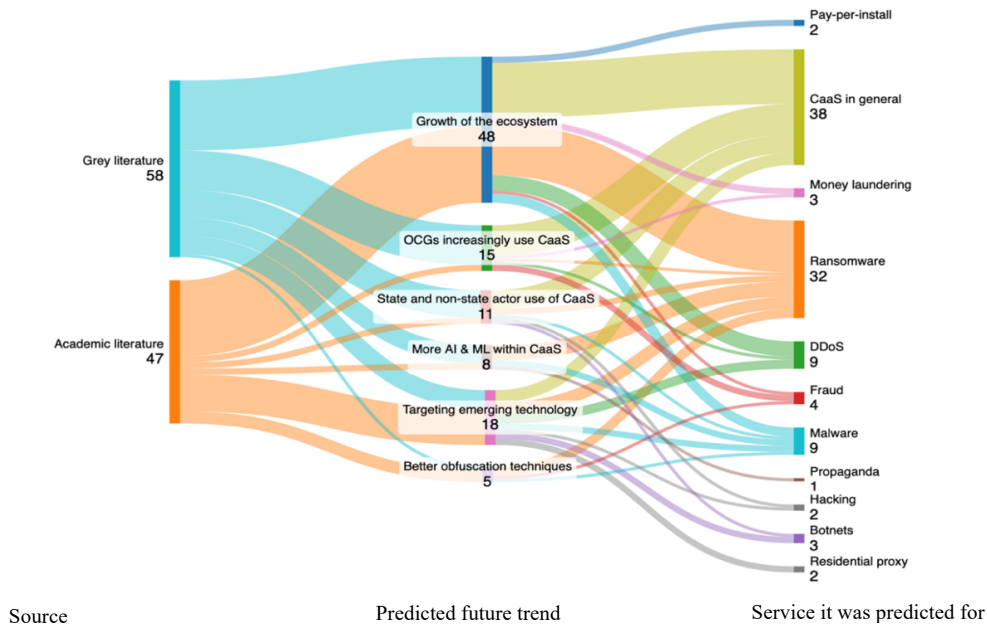

Figure 7: Future trends predicted for different CaaS services

17 articles included in this review presented CaaS as one of the emerging tools used by organised crime groups (OCGs). This involves making use of CaaS services such as Money Laundering and (Identity) Fraud as a Service in addition to potentially discovering new avenues of revenue with engagement in online criminal activities despite offenders lacking such expertise [192]. Money laundering, in particular, is intricately linked to nearly all criminal activities both in the cyber and physical realm and is provided as a service by highly specialised criminal groups [196]. The literature is also points towards evolution the use of privacy-centric cryptocurrencies across the CaaS ecosystem. Groups have been noted to be shifting from Bitcoin to Monero, TRON, and Tether to avoid crypto tracing during fundraising, laundering and purchasing

CaaS tools [206]. CaaS providers are also increasingly using cross-platform, cross-chain bridges, and coin-swapping services to move funds across blockchains. This complicates tracking and enhances anonymity further [206].

Table 5: Future trends as discussed in the literature

| Future trend | Description | Source | Papers | Count |
|---|---|---|---|---|
| Growth of the business model | CaaS model predicted to grow in the number of services provided and further commercialisation and move towards a legitimate business model. | Academia | [137], [68], [105], [79], [99], [37], [108], [121], [93], [118], [122], [248], [107], [116], [124], [142], [64], [123], [81], [98], [139], [132], [30], [70], [15], [138], [144], [88], [14], [106], [200], [94], [95], [96] | **34** |
| | | Grey literature | [176], [189], [147], [191], [161], [208], [176], [151], [225], [28], [175], [219], [241], [146], [150], [186], [223], [172], [188], [148], [216], [183], [212], [224], [187], [185], [234], [58], [204], [201], [192], [185], [57], [227], [205], [206] | **36** |
| CaaS increasingly used by organised crime groups (OCGs) | Increased use of CaaS-enabled services by OCGs. This may span cyber-enabled activities such as fraud, money laundering and identity theft to cybercrime attacks. | Academia | [99], [248], [70] | **3** |
| | | Grey literature | [243], [164], [249], [179], [160], [152], [147], [162], [161], [151], [150], [172], [187], [196] | **14** |
| CaaS increasingly used by state and non-state actors. Targets may include CNI | CaaS products may be used in the context of cyber war, cyber terrorism and cyber espionage. These groups may target critical national infrastructure. | Academia | [98], [132], [14], [95] | **4** |
| | | Grey literature | [179], [170], [165], [236], [175], [146], [242], [148], [234], [200], [57], [229], [239], [232], [208] | **15** |
| Abusing/targeting emerging technology | Services under the CaaS model may target IoT devices, cloud devices, CAVs and so on. | Academia | [93], [118], [122], [107], [82], [124], [123], [139], [138], [94], [129], [96], [95] | **13** |
| | | Grey literature | [176], [237], [242], [173], [57] | **5** |
| More sophisticated obfuscation techniques | Antivirus programs and technologies such as endpoint detection and response already have difficulties detecting malware. This is likely to increase due to tougher competition and more services specialising in obfuscation alone. | Academia | [60], [61], [116], [81], [139], [144], [68] | **7** |
| | | Grey literature | [210], [189], [28], [185], [206] | **5** |
| Increased use of AI and ML in services provided | Use of AI and machine learning (ML) both as malicious, standalone services such as deep fakes are entwined with other services to improve their operations. | Academia | [250], [61], [118], [141], [62], [106] | **6** |
| | | Grey literature | [207], [237], [28], [216], [173], [216], [192], [56], [193], [57], [206] | **11** |
| Move to social media | CaaS model to move from dark web forums to social media platforms. | Grey literature | [208], [206] | **2** |
| Privacy at risk | More CaaS services gathering personal information online in an unregulated manner which may cause privacy concerns. | Academia | [250] | **1** |
| | | Grey literature | [237] | **1** |

Similarly, larger organisations such as state and non-state actors and extremist groups were identified as increasingly commonly becoming customers of the CaaS ecosystem [239]. CaaS products may be used in the context of cyber war, cyber terrorism and cyber espionage [14, 200, 208]. The line between financially motivated and ideologically motivated actors is blurring as hacktivists and extremists may increasingly use CaaS tools for political or social disruption [206]. Sources are pointing towards the increasing intersection between cybercrime and cyber warfare [229]. In particular, these groups may target critical national infrastructure (CNI) facilities which would have catastrophic consequences. Combining a physical attack with a cyber one to create a hybrid attack may amplify the damage caused exponentially [57]. CNI predicted to be at future risk may span from electric grid systems (see for example [239]), oil and gas [216, 231], finance sector [94] and even hospitals [232]. All of these have varied cyber capacities which means that it is difficult to implement overarching cyber standards to ensure secure operations. Furthermore, CNI facilities have numerous partner organisations and interconnected systems that may present vulnerabilities which may lead to large monetary losses or even the loss of lives. The services mentioned in these circumstances were often Ransomware and other forms of Malware as a Service, DDoS, Hacking and Botnets as a Service. These may be used either for gathering financial support for organisations, cyber-attacks or hybrid attacks.

As with all threats within the field of cybersecurity, the arms race between malicious actors and security providers is constantly pushing the boundaries of innovation. As new technologies continue to come online, the attack surface continues to increase too. This includes an increased targeting of emerging technologies such as Internet connected or Internet of Things (IoT) devices, cloud services, Connected and Autonomous Vehicles (CAVs) and cryptocurrencies to name a few. In the literature, authors noted that not only will these technologies become new targets, but some of the CaaS services may rely on them to succeed. Examples include the use of cryptocurrencies for profit transactions and using IoT devices within larger botnet structures [57]. This, in turn, amplifies the attack as well as creates more profitable cybercriminal infrastructure, in this case, botnet infrastructure. Numerous papers suggested that the cloud will be a major target in the coming years. By targeting environments such as the cloud, malicious actors may achieve larger supply chain compromises and even manipulate stock markets [94, 96]. In terms of targeting emerging technologies, the services discussed specifically were residential proxy, ransomware, DDoS, botnets and ransomware. However, it may be safe to assume that any services within the model may attempt to take advantage of emerging technology.

The CaaS model is also predicted to improve the obfuscation strategies cybercriminals use to make their services stealthier and harder to detect by both simply having more roles in the ecosystem focusing on better obfuscation but also increasing the use of smart and self-evolving services. These may continue to exploit multi-factor authentication and cybersecurity products such as Endpoint Detection and Response technologies [185]. This can be achieved by including some level of artificial intelligence within the service, such that it evades most forms of static and dynamic detection attempts. Sophisticated obfuscation may also be achieved by increased creativity of the delivery of malware to the system, such as embedding malicious text file containing a series of PowerShell commands that automate tasks and manage system configurations into digital images [116]. Another part of the cybercrime ecosystem to which increased obfuscation pertains concerns payment demands where malicious actors are using more privacy-preserving cryptocurrencies (e.g. Monero) when demanding ransom [68].

The final larger trend predicted in the literature refers to the increased use of artificial intelligence and machine learning within the services themselves, not just for the purposes of obfuscation, but to facilitate and improve attacks and maximise opportunities for profit. There are many instances of this including the creation of deepfakes. Artificial intelligence-enabled predictive language tools and chatbots have made it easier for cybercriminals to develop advanced forms of ransomware or even create independent web crawlers that collect personal data that is then used further in the ecosystem. Both Malware and RaaS are predicted to include more AI and ML capabilities as well as increase the stealthiness present in operations. The use of (gen) AI may both introduce new attack vectors as well as ease up the administrative load it takes to run a Cybercrime operation [57]. In the case of Fraud and Disinformation as a Service, authors predict that (gen) AI will also play a big role in social engineering as technologies such as Deepfakes may be readily available on the market [193].

## 5 DISCUSSION

Cybercrime as a Service is not a new phenomenon. It has featured prominently in the yearly Internet Organised Crime Threat Assessments produced by Europol since 2014 onwards. However, to the best of our knowledge this is the first scoping literature review on the topic.

One of the main observations of the review (which was not an original research question) is that CaaS is often spoken about as an accessory to other cyber threats rather than a core focus of the literature analysed. In the grey literature, we observed it in every report reviewed. In these reports, the authors focussed mainly on threat actors being involved in a range of different types of cybercrime, how they work and the implication of CaaS for the different types of offending considered. Relative to the academic literature, grey literature tended to have a wider scope and discussed CaaS to a more limited extent. In the academic literature, there was a subset of work that prioritised the investigation of the CaaS model and its ecosystem (mainly work in the categorisation of evidence I, II, and partially V), however, the largest group (Categorisation III), focussed on a very narrow and often technical problem such as the detection of a particular type of malware. In such papers, CaaS was only mentioned as a potential enabler of that threat and was not a central aspect of the research. Given the potential impacts of CaaS, this general finding that it was rarely the focus of published work was somewhat unexpected.

The first research aim of this paper was to study what services have been the focus of published research. In addition to the papers that spoke broadly about the CaaS model and its ecosystem, the most common services reviewed were Ransomware, Malware, and Distributed Denial of Service. These were followed by illicit services that enable other malicious activities such as Botnets, Phishing and Money Laundering as a Service. Some of the newer attack vectors included Concession abuse, Spyware, and Stealers as a Service.

Another important finding of the review relating to the services studied was the modularity of CaaS services and the lack of standardisation among the services presented in the literature. Different authors draw the line amongst, or define what a particular service is, differently. In other words, there is a lack of clarity regarding the scope of the services described by authors and their interconnections with other services. For example, when authors discuss Botnet as a Service, they typically do not explain how the malware infects machines to become part of the botnet (which could be part of a Phishing as a Service campaign, installed by Malware as a Service, etc.), the Command and Control infrastructure of the botnet (is this a separate service or part of the botnet's infrastructure?), the attacks carried out by the botnet (such as RaaS or DDoS as a Service), or how the illicit proceeds are funnelled to the malicious actors (often involving Money Laundering as a Service). Other authors may discuss each (or a smaller number) of these as a distinct service, without tying them together in a single problem space.

Establishing a framework with clearer definitions is crucial for aiding law enforcement, security professionals, and everyone affected by CaaS to categorise, measure crimes, and propose effective responses and prevention strategies [251]. Additionally, knowing that what we measure is consistent with other research ensures the reliability of the measurements. This would also help policy makers involved in cost benefit analyses of prevention strategies [256]. Conceptual clarity is, of course, also important for academia where authors across scientific fields highlight that scientific progress depends on the establishment of a shared theoretical framework, which defines the core concepts and research agendas of a discipline [252-254]. Moreover, while the concept of CaaS has gained traction in both grey and academic literature, its scientific utility in explaining the evolution and performance of the cybercrime ecosystem remains under explored and underutilised. The term is often used descriptively as a factor enabling a given cybercrime trend rather than analytical papers focussing precisely the ecosystem itself and what propels it to be so widely spread. Where more detailed analysis is conducted, the papers feel quite disjointed from the rest of the literature, thus restricting its ability to support longitudinal or comparative studies. Without a shared understanding of what constitutes a malicious "service", how they are offered, or how services interact within the ecosystem and each other, it becomes difficult to assess how cybercrime operations scale, adapt, or specialise over time. This lack of conceptual clarity also hampers efforts to measure the efficiency, resilience, and innovation within cybercriminal networks. A further research gap identified is the mapping of current and future enabling technologies that CaaS operators and service providers use. This should be done, using a more robust theoretical framework, defining service boundaries, actor roles, and ecosystem dynamics, to enhance understanding of the CaaS ecosystem.

A potential way to increase standardization would be to use a framework that builds on the Cyber Kill chain [255] to understand CaaS and its various modular Services. The Cyber Kill Chain is a widely used framework developed by Lockheed Martin to analyse the stages of cyber-attacks and help organizations enhance their defences. It consists of seven steps, each representing a phase of the attack lifecycle, from intelligence gathering to data exfiltration or system compromise [256]. Since its creation, the Cyber Kill Chain has influenced cybersecurity practices by offering a structured approach to both detect and prevent attacks more effectively [72, 257, 258]. This framework extends well to CaaS, giving the opportunity of intervening against CaaS actors on a technical point of view as much as adding another way to describe the phenomena to the holistic approach that is necessary in this field. The Cyber Kill Chain does not only highlight technical countermeasures but identifies a set of stages in the CaaS pipeline that could be analysed using techniques such as crime scripts [259], which are used in criminological research to describe the steps offenders engage in to commit offences.

Building on from the lack of definitions, as CaaS has implications for so many sectors and stakeholders, another issue is the standardisation of language and the use of the "as a Service" denomination. For example, in a grey literature article on the next generation of Organised Crime [260], the authors discussed the implications of emerging trends for the organised crime business model in depth, but did not use the term "as a Service" once. As a result, this paper was not included in our study. Similarly, we found that many authors used unspecific expressions such as "criminal services", "cybercriminal syndicates" (see for example [200]) or "cybercrime services being offered over dark web" (see for example [55]) which may hint towards criminal services being offered "as a Service" without explicitly saying this. This makes it difficult to locate potentially relevant studies using search strategies such as that used here. Considering this issue more broadly, it is possible to say that there is no shared terminology, sometimes even between authors within the same field. While this may be due to the still young and evolving landscape, challenges related to the absence of a shared vocabulary and taxonomy can hinder progress. A future research priority then, would be to build consensus on a taxonomy through consultation with researchers from different disciplines as well as the wider stakeholder community.

Part of our inclusion criteria was that studies had to focus on services that are created with malicious intentions in mind. When screening the articles, we noticed that the literature may have missed some potential threats. For example, the majority of services under the XaaS model such as Software, Infrastructure, and Function as a Service are not malicious, however, this does not mean they cannot be (and are not) misused. Similarly, Machine Learning as a Service has become widely adopted for many applications in recent years, and multiple papers discuss associated vulnerabilities such as data poisoning where a malicious actor injects fake training data to corrupt a model, rendering the service useless or even malicious. This is especially important in the context of cybersecurity as machine-learning approaches are increasingly used within the Security as a Service paradigm to identify potential vulnerabilities in a given system so that cybersecurity teams can patch them. However, there is nothing stopping a threat actor from using similar tools to help to develop and/or deploy an attack. However, since that was not the intent of the service from the start, it arguably does not belong in CaaS category and would not have been covered in this review.

Furthermore, dual-use technology was also beyond the scope of this project, but the authors believe that this should be investigated further, as research often overlooked the vulnerabilities, misuse, and security aspects. A number of articles were excluded in the screening process, as they were discussing Drones as a Service technology from a non-malicious perspective. Examples of the papers that were excluded discussed assisting in search and rescue missions, but they failed to mention that those technologies can also be used as weapons.

The literature suggests that we are to expect organisations commodifying crime to develop an even more professional service environment. Thus, moving from a single cybercrime provider with fewer criminals operating it (and having a single point of failure) to a larger division of labour with more specialism, competitiveness, and resilience. This may include further inclusion of more specific roles within the ecosystem, that work as separate entities and cooperate with one another, as any business would. With regards to actors participating in CaaS, the lowered barrier to entry and the global recession on the horizon, CaaS may offer a new mode of "full-time work" to multiple new actors, regardless of their technical ability [185]. Avgetidis et al. [53] examined over 4500 stealer operators and their patterns of behaviours that occured over the course of a 24-hour period with a diurnal analysis and the findings suggest that operators administer their stealer malware as a full-time job with higher activities during working hours of weekdays and almost no activity over weekends. Furthermore, the work by Ruellan et al. [67] investigated internal communications within the Conti ransomware group. The authors highlight that the Conti ransomware group provided customer support to its affiliates (the "customers") through dedicated channels, which is a key aspect of their operations. This support structure was professional and well-organized, including assistance with technical issues related to ransomware deployment, payment processing, and negotiation strategies with victims. The presence of a structured customer support team reflects the sophistication of the Conti RaaS operation, which functioned more like a legitimate business, offering comprehensive services to its affiliates to ensure smooth execution of attacks and maximize profits. Following this example, as the business develops and grows financially, the market will grow, and attacks will grow in their sophistication to keep the competitive advantage. With our lives moving online and cities and critical national infrastructure moving towards smarter and more cyber-physical systems, attacking these systems may have even more significant consequences on the public, unless secured properly.

Numerous authors (e.g. [14, 37, 68, 94, 105, 107, 108, 116, 124, 200, 261, 262]) discussed the fact that for CaaS to grow as a business model there will be a need for advertising and services to be increasingly accessible. Thus, it has been predicted that there may be a shift in distribution and advertising from dark web forums to x\more generic platforms such as Discord and Telegram. Since the actual transactions, and customer support provided after a service has been procured, typically take place on private communication channels on Telegram, Discord, Wicker, and Jabber [263, 264], it may not be too long before advertisement and the rest of the supply chain also moves to these platforms. Such platforms have already become increasingly popular sources of communication amongst criminals since the takedowns of dark web markets including AlphaBay and Hansa in 2018 [265]. There are multiple reasons for this including the privacy provided through end-to-end encryption, the fact that such channels bypass market fees, and because they provide a larger and more easily accessible customer pool (especially non-technical customers which are arguably the target audience of CaaS) [206]. In addition, a great deal of administrative work is already taken care of by the social media provider, as opposed to offenders running their own forums [7]. However, to continue to operate, and avoid being banned, operators and advertisers using these platforms will, of course, have to evade the illicit content detection algorithms used by platforms. It is worth noting that many of the experts in the field have warned about the presence of CaaS on social media at scale, including Signal and Telegram. However, apart from two grey literature articles, this was not reflected in our literature sample, making this a research gap that needs to be addressed.

This increased accessibility may further open the doors for more malicious actors to become customers of CaaS and even offer services themselves. This may include OCGs and extremist actors. The outsourcing of activities that these groups may previously have lacked expertise in might change their organisational structure to more business-like structures [172] [187]. With respect to (non)state actors, CaaS has already been used in cyber-warfare. For example, there have been multiple reports of CaaS models such as ransomware, malware and DDoS as a Service operations being used in the Russia-Ukrainian conflict [266, 267]. As CaaS use may increase, so will the issues associated with the attribution of attacks [186].

## 5.1 Limitations and future work

As with any research, this review is not without limitations. Firstly, grey literature was included within the scope of this scoping review, however, the literature included primarily consisted of government and policy papers. We realise that there is an active research community on this topic within the industry sector and as such, future research may consider analysing trends published within trade journals, white papers and industry reports. However, the decision to exclude industry reports was twofold – to our knowledge, there is no systematic way of gathering industry reports and industry reports may be biased as they are partially written for commercial purposes.

The second limitation concerns the lack of standardisation in terms of terminology and approaches within different disciplines. As this scoping review draws on findings from multiple disciplines this is important, and the authors had to make numerous choices when determining the inclusion criteria. For example, we had to decide where to draw the line when defining 'cyber-enabled services,' and how to differentiate whether a service is inherently malicious or simply has the potential for misuse which is often a grey area. Furthermore, fields like criminology and sociology may be studying topics related to online cybercrime marketplaces and their evolutions on the clear and dark web, however, as their terminology may not include CaaS or "for hire" such papers may not have been identified in our search. A potential future direction would be to generate some sort of topological approach in order to standardise the everchanging field of CaaS, expand the inclusion criteria to include insights from disciplines using different terminologies, and thus ensure further systematic evaluation of the issue.

## 5.2 Implications

Historically, digital undergrounds typically consisted of actors who were technically less experienced, such as script kiddies. Such actors presented a lower threat in terms of carrying out comprehensive cyber-attacks. However, with CaaS this no longer holds true [2]. As such, Europol have remarked that CaaS must become a top priority for law enforcement globally [2]. Moreover, CaaS acts as a multiplier for many facets of cybercrime - services such as illegal currency exchanges, money mules, and bulletproof hosting services that may not be the direct objective of CaaS but are crucial for the economy's upkeep. This may also have implications for both cybercrime as well as urban acquisitive crimes as, for example, organised crime members may use those services to expand their operations or launder the proceeds of predicate offences [70, 248].

Due to the increasingly complicated and commercialised ecosystem, the community needs to study CaaS and its ecosystem in a standardised way. One approach would be to apply the cyber kill chain as described above. However, arguably, the most effective way to counter the CaaS ecosystem is to disrupt the infrastructure that enables it, as this addresses the roots of cybercriminal activity rather than just the symptoms. CaaS networks and enablers rely heavily on a robust infrastructure of servers, networks, forums, command and control infrastructure and malicious software, which facilitates everything from data theft, malware deployment, and anonymous communication [11]. By dismantling or compromising these elements, such as through botnet takedowns or server seizures, authorities can cripple cybercriminal operations at a foundational level, reducing the immediate and long-term threat [268]. This approach also discourages reinvestment in expensive systems that support these activities, promoting a more sustainable path to cybersecurity.

Due to the complicated network of actors involved in CaaS, some sources are warning practitioners against investing resources into attack attributions, as attempting to do so will become increasingly complicated [269]. As the service provider (i.e., the original creator of the cyberattack) is further removed from the end attack, identifying and eliminating the original source becomes a significant challenge. This may be the only sustainable way to dismantle the service, but it is not impossible. There have been notable arrests in the field, such as Russian authorities raiding one of the largest RaaS operators, REvil, in late 2020. They arrested 14 of its members and halted its operations at the request of U.S. officials [29]. More recently, in May 2024, Operation Endgame disrupted services in the botnet and dropper malware ecosystems [270]. This had a longer lasting effect as the dropper malware was used for delivery across many services.

# 6 CONCLUSION

Cybercrime as a Service is a growing economic model where technically skilled actors offer toolkits as a packaged service that, for a fee, enable less experienced bad actors to carry out successful cyberattacks. There are numerous malicious services under this model which continue to expand and increasingly take the form of a business-like structure, making the eco-system resilient, scalable and accessible by more and more actors. Consequently, CaaS is lowering the barrier of entry to cybercrime, the services are becoming more sophisticated and harder to detect and remove.

The aim of this study was to take stock of what services have been studied in the literature, what methodologies have been employed in those studies, what gaps exist, and how CaaS might evolve in the future. To this end, we conducted a scoping review of the literature initially consisting of 2434 articles, collected from six academic and two grey literature databases, with 195 articles being included in the final, content analysis.

With 15 Services being studied in the academic literature and 19 in grey literature, we can see that the CaaS research is diverse. A variety of methodological approaches from different disciplines were applied – from studying detection capabilities, reverse engineering CaaS attacks, risk assessment, case studies and response analyses and variety of narrative literature reviews examining CaaS and its consequences. Often CaaS would not be the focus of the projects, which may explain why there is such a variety of nomenclature and dispersion of research findings. When narrowing down on papers with a focus on CaaS, there was 2.5 times more quantitative papers than qualitative. Thus, some of the research gaps on the topic from a methodological perspective, the research community should address are systematic analysis of case studies, law enforcement response and consultation of experts surrounding the CaaS ecosystem. Furthermore, there were only a handful of studies investigating CaaS listings in their original marketplace (dark web forums or elsewhere), so further research efforts should also address this. Market place analysis may have been conducted in a variety of fields (e.g. criminology, anthropology and sociology), however, if scholars in those fields. may not frame malicious transactions and emerging cybercrime organisational structures through the lens of CaaS, creating a barrier to the integration of these literatures. Going forwards, only by developing a holistic, structured approach to gather insights across disciplines, can the phenomena be understood and effective approaches to prevention be developed. Furthermore, there has only been one economics paper included in the review, looking at CaaS through a game theory lens. As CaaS is becoming more of an established business, approaching the field as such and applying more economic theories might help us to understand and predict the changes of the market better.

Further development of the CaaS business model and its move to more accessible platforms such as social media may increase the number of its customers. Without further research and action taken by decision-makers, the model will likely be increasingly used by organised crime groups, terrorists and larger state actors, whose capabilities may result in large-scale consequences affecting many civilians. It is evident from this work that multistakeholder cooperation will be vital in curbing the impact that CaaS may have on both cyber and physical crime worldwide. Further efforts have to be made into understanding the operational aspects of CaaS, its enablers and the underlying infrastructure in order to develop sustainable countermeasures. Only a holistic and systematic approach to the topic will allow the community to understand how the threat of Cybercrime as a Service will develop going forwards and then take steps to disrupt it.

**ACKNOWLEDGMENTS**

This project was funded by the UK EPSRC grant [EP/S022503/1] that sup- ports the Centre for Doctoral Training in Cybersecurity delivered by UCL's Departments of Computer Science, Security and Crime Science, and Science, Technology, Engineering, and Public Policy.